\documentclass[journal=nalefd,manuscript=article]{achemso}

\usepackage{amsmath}
\usepackage{amssymb}
\usepackage{gensymb}
\usepackage{siunitx}
\usepackage{xcolor}
\usepackage[utf8]{inputenc}
\setkeys{Gin}{draft}

\author{Artur Avdizhiyan}
\affiliation{Faculty of Physics, University of Bialystok, 15-245 Bialystok, Poland}
\author{Weronika Janus}
\author{Marcin Szpytma}
\author{Tomasz Ślezak}
\affiliation{Faculty of Physics and Applied Computer Science, AGH University of Krakow, Kraków, Poland}
\author{Marek Przybylski}
\author{Maciej Chrobak}
\affiliation{Academic Centre for Materials and Nanotechnology, AGH University of Krakow, Kraków, Poland}
\altaffiliation{Faculty of Physics and Applied Computer Science, AGH University of Krakow, Kraków, Poland}
\author{Vladimir Roddatis}
\affiliation{Helmholtz Centre Potsdam, Telegrafenberg, D-14473 Potsdam, Germany}
\author{Andrzej Stupakiewicz}
\author{Ilya Razdolski}
\email{i.razdolski@uwb.edu.pl}
\affiliation{Faculty of Physics, University of Bialystok, 15-245 Bialystok, Poland}

\title[]
  {Ultrafast laser-induced dynamics of non-equilibrium electron spill-out in nanoplasmonic bilayers}


\begin{document}

\begin{tocentry}

The surrounding frame is 9\,cm by 3.5\,cm, which is the maximum
permitted for  \emph{Journal of the American Chemical Society}
graphical table of content entries. The box will not resize if the
content is too big: instead it will overflow the edge of the box.

\end{tocentry}

\begin{abstract}
Contemporary quantum plasmonics captures subtle corrections to the properties of plasmonic nanoobjects in the equilibrium. Here, we demonstrate non-equiulbrium spill-out redistribution of the electronic density at the ultrafast timescale. As revealed by time-resolved 2D spectroscopy of nanoplasmonic Fe/Au bilayers, an injection of the laser-excited non-thermal electrons induces transient electron spill-out thus changing the plasma frequency. The response of the local electronic density switches the electronic density behaviour from spill-in to strong (an order of magnitude larger) spill-out at the femtosecond timescale. The superdiffusive transport of hot electrons and the lack of a direct laser heating indicate significantly non-thermal origin of the underlying physics. Our results demonstrate an ultrafast and non-thermal way to control surface plasmon dispersion through transient variations of the spatial electron distribution at the nanoscale. These findings expand quantum plasmonics into previously unexplored directions by introducing ultrashort timescales in the non-equilibrium electronic systems. 
\end{abstract}


Fast and efficient control of collective electronic excitations remains one of the main challenges of active plasmonics \cite{MacDonaldNatPhot2009,MetzgerACSPhot2016,KimNanop2022}. Notably, classical approaches to plasmonics start to break down when the dimensions of plasmonic systems enter the nanometer scale, which has recently stimulated the explosive growth of quantum plasmonics with rich underlying physics and intriguing novel phenomena\cite{TameNatPhys2013,MarinicaSci2015,FitzgeraldIEEE16,ChristensenPRL17,DingJPCM18}. Incorporating non-locality of the optical response and non-classical plasmon damping mechanisms, novel theoretical approaches have been developed and intensely debated \cite{MortensenNanoph2021,StamatopoulouOME2022}.

\begin{figure}[t]
  \includegraphics[width=8cm]{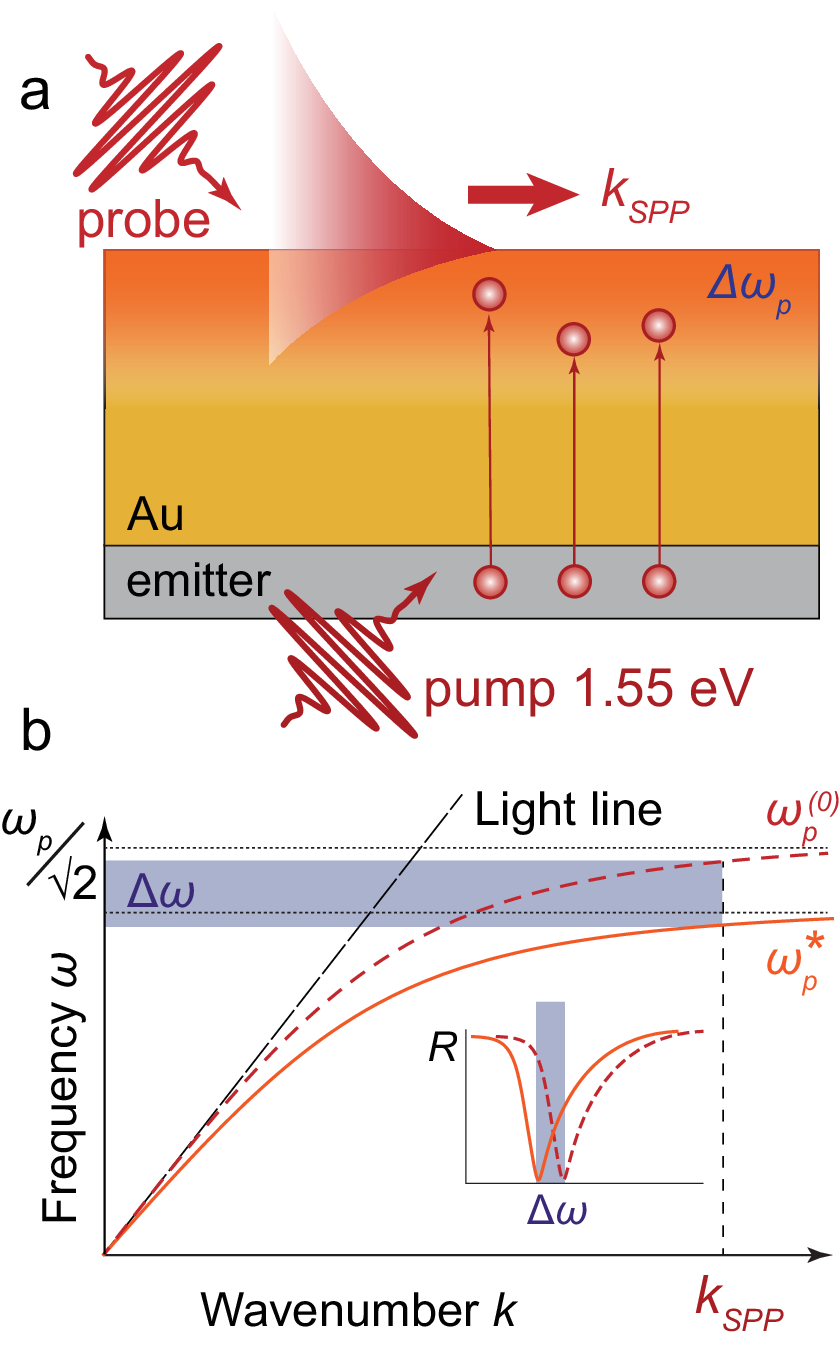}
  \caption{The concept of non-thermal electron-induced remote control of the SPP dispersion. (a) Schematics of the experimental approach. Laser-excited hot electrons in the emitter traverse the Au layer in the superdiffusive regime. Arriving at the Au surface, they introduce variations to the plasma frequency $\Delta\omega_p$. 
  (b) The SPP dispersion $\omega(k)$ (orange solid line) at a new, non-equilibrium plasma frequency $\omega_p^*$ is displaced with respect to the equilibrium case ($\omega_p^{(0)}$, red dashed line), observed as a SPP resonance shift $\Delta\omega$ in the experiment.
  }
  \label{fig:1}
\end{figure}

One of the most intriguing quantum effects at the nanoscale pertains to the spill-out of the electronic density $n_e$ when the wavefunctions of the conduction electrons surmount the material boundaries \cite{Zhu2016}. While tackled by multiple approaches,  the electron spill-out is satisfactorily accounted for within the Feibelman formalism \cite{Feibelman76,Feibelman82} employing two effective parameters $d_{\parallel}$ and $d_{\perp}$. {\color{red} In particular, this approach has been developed and utilized to account (in a semi-phenomenological way) for the non-locality which as a fundamental physical concept plays an essential role at the nanoscale. Instead, a local dielectric function $\varepsilon$ is used while the electronic density of the metal acquires an additional effective contribution localized at the interface. Termed as a surface-response formalism \cite{Feibelman82}, this approach is computationally advantageous and can be employed to incorporate quantum effects (Landau damping, spill-out/in etc.) into an essentially classical description \cite{Feibelman82,LiebschPRB87}. It is further instructive that the Feibelman parameters  correspond to the the centroid of the induced charge density and of the normal derivative of the tangential current density, respectively.}
Depending on the geometry, quantum plasmonic effects in noble metals are usually well described with $d_{\parallel}=0$ and $|d_{\perp}|\lesssim10^{-1}$~nm \cite{ChristensenPRL17,RodriguezEcharri21}. Introduced in this way, $d_{\perp}$ can be related \cite{Apell81} to the variations of the plasma frequency $\omega_p^2=4\pi n_e e^2/m$, where $m$ and $e$ are the electron mass and charge, respectively:

\begin{equation}
    \omega_p\approx\omega_p^{(0)}(1+\frac{1}{2}kd_{\perp})
    \label{eq:feibelman}
\end{equation}
Here $\omega_p^0$ is the unperturbed plasma frequency, and $k$ is the in-plane wavenumber \footnote{Derived for $\varepsilon_{\infty}=1$. An interband contribution $\varepsilon_{\infty}$ in noble metals results in a small correction to $d_{\perp}$.}. Owing to the large optical wavelength, this Feibelman correction to $\omega_p$ is usually on the order of $10^{-3}$ or smaller. In quantum plasmonics, this effect induces a dispersion shift of a surface plasmon-polariton (SPP) mode. Indeed, the SPP dispersion $\omega(k_{\rm SPP})$ can be unambiguously related to $\omega_p$, since in the $k_{\rm SPP}\rightarrow\infty$ limit, $\omega_{\rm SPP}\rightarrow\omega_p/\sqrt{2}$ at the metal/air interface. As such, any perturbation to $\omega_p$ results in a shift of the SPP resonance (cf. Fig.~\ref{fig:1}b).

Interestingly, this approach has been developed for the thermal equilibrium, where the electronic distribution is essentially Fermi-Dirac. Yet, significant modulation of the plasmonic properties can be expected in strongly non-equilibrium conditions featuring non-thermal electrons. Owing to their short lifetimes, femtosecond stimuli are required, allowing for studying the dynamics of surface plasmons at their intrinsic timescales \cite{BaidaPRL2011,BruggemanPRB2012,PohlPRB2013,Raschke,FischerOptica2021}. A promising method for the generation of a non-equilibrium electron population consists in the laser-induced non-thermal electron injection. Experimentally, an emitter layer exposed to the laser pulse acts as a source of highly energetic laser-excited electrons (Fig.~\ref{fig:1}a). Observed in Fe/Au bilayers and termed as the non-equilibrium Seebeck effect, it results in the generation of a current pulse propagating away from the excitation spot. The subpicosecond duration of this pulse is governed by the lifetime of the non-thermal electrons in the Fe emitter\cite{MelnikovPRL11,AlekhinPRL17}. Notably, the injected electrons can propagate over large ($\sim10^2$~nm) distances in the superdiffusive regime \cite{BrorsonPRL87,LiuPRB2005,MelnikovPRL11,AlekhinJPCM2019,MelnikovPRB22,KuhnePRR2022}, thus decoupling the direct laser irradiation from the eventual plasmonic excitation. Bringing the perturbation to the remote plasmonic interface, non-thermal electrons modify $\omega_p$ and thus the SPP dispersion at the ultrafast timescale (Fig.~\ref{fig:1}b). The latter minimizes the effect of heating of the plasmonic medium, which is known to primarily enhance the losses\cite{AlabastriMater2013,ReddyOPM16, SchiratoNanoLett2022}.

\begin{figure}[t]
  \includegraphics[width=8cm]{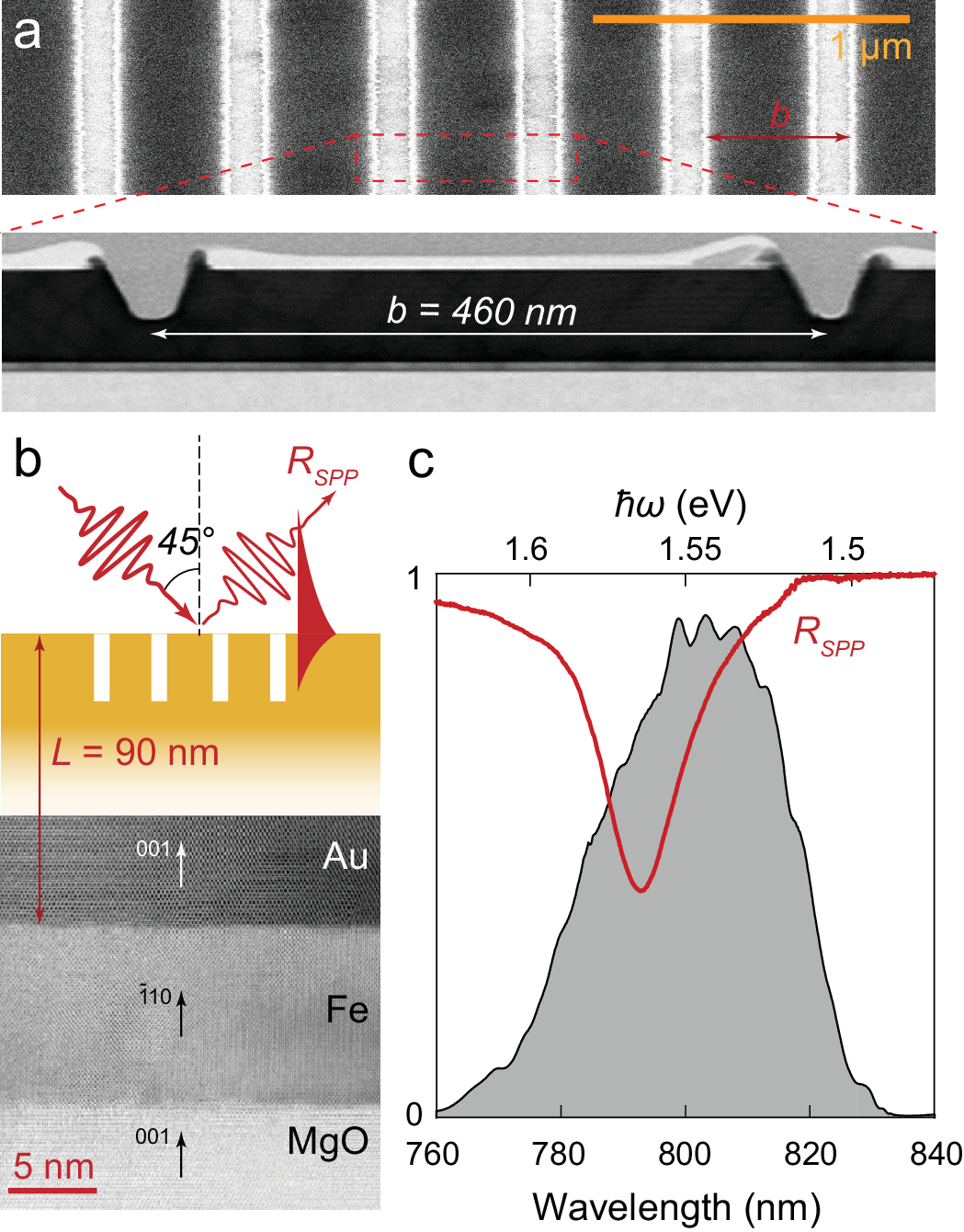}
  \caption{(a) Optical microscopy image of the periodically corrugated surface (top) and bright field scanning transmission electron microscopy (STEM) image of the cross-section (bottom). The deep trenches in Au are visible, forming the periodic grating. (b) The epitaxial Au/Fe/MgO sample: structure, experimental configuration, and high-resolution STEM cross-section image illustrating high quality of the interfaces. (c) Optical reflectivity spectrum measured on the periodically corrugated Au surface. The dip at 793~nm indicates the SPP excitation. The grey shaded area is the spectrum of the 60 fs-long incident probe pulse.}
  \label{fig:2}
\end{figure}

In this work, we investigate ultrafast dynamics of the hot electron-driven quantum plasmonic effects at the femtosecond timescale. To that end, we perform time- and wavelength-resolved spectroscopy of the SPP resonance in strongly non-equilibrium conditions. Exciting an epitaxial Fe/Au bilayer (Fig.~\ref{fig:2}), laser pump pulses (60 fs duration, 1.55 eV photon energy, 1~kHz repetition rate) are focused into a spot of $80~\mu$m in diameter. The resulted laser fluence on the order of 40~mJ/cm$^2$ is kept below the damage threshold of the Fe film. The pump pulse energy absorbed in Fe induces the non-thermal electron injection into Au. While these electrons propagate towards the outer Au surface, introducing a non-equilibrium state into the local electron density, we monitor it response through the dispersion dynamics of the SPP mode. The latter is excited by delayed, weak ($>30:1$ pump-probe energy ratio) probe pulses impinging at $45\degree$ at the periodically corrugated Au surface (Fig.~\ref{fig:2}). The reflected from the sample probe beam is sent into an HRS-300 monochromator and then registered by a spectroscopic BLAZE CCD camera (both by Princeton Instruments) synchronized with the laser pulses. We register the pump-induced differential reflectivity of the probe spectrum $\Delta R/R$~$(\lambda,t)$ around 800 nm central wavelength with the femtosecond temporal and sub-nanometer spectral resolution. 

An MgO(001) $10\times10$~mm$^2$ square substrate was annealed at 780~K before the deposition. Then, a 4 nm-thin homoepitaxial MgO buffer layer was deposited at 725~K using electron beam evaporation. This procedure ensures high epitaxial quality of the MgO surface and helps to avoid contamination. Next, a Fe(10 nm)/Au(90 nm) bilayer was grown on top of the MgO surface at room temperature under ultrahigh vacuum conditions with molecular beam epitaxy. The high-quality epitaxial growth of all layers was confirmed by the low-energy electron diffraction (LEED). The LEED pattern (not shown) corresponding to the Fe(001) surface indicated the expected $45\degree$ in-plane rotation of the Fe lattice with respect to that of MgO, so that Fe[110]//MgO[100]. Similarly, a $5\times1$-type reconstruction of the Au(001) surface together with the $45\degree$ in-plane lattice rotation with respect to Fe(001) can be concluded from the Au(001) LEED pattern. Finally, an excellent epitaxial quality of the interfaces (Fig.~\ref{fig:2}) was confirmed by the cross-section transmission electron microscopy (Themis Z, Thermo Fisher Scientific, equipped with a corrector of spherical aberration at the probe side) operated at 300 kV. 

To enable SPP coupling, the Au(001) surface was then laterally patterned by electron lithography with a positive resist. Carried out with Ar-ions of 1~kV energy, the etching produced 35~nm-deep nanotrenches. The pressure in the vacuum chamber during the etching was on the order of $9\times10^{-6}$~mbar and the current density was $15$~$\mu$A/cm$^2$. In order to avoid overheating of the sample, the process was divided into three subsequent steps: 18 minutes of etching, a 30-minutes break and another 18 minutes of etching. The architecture of the resulted Au surface demonstrated a periodic structure of 500~$\mu$m-long nanolines parallel to the Fe[110] direction and separated by 70~nm-wide trenches (Fig.~\ref{fig:2}). The $460$~nm periodicity enabled the excitation of a quasi-SPP mode at the Au/air interface with the incident probe beam in the vicinity of a $800$~nm wavelength.

\begin{figure}[t]
  \includegraphics[width=8cm]{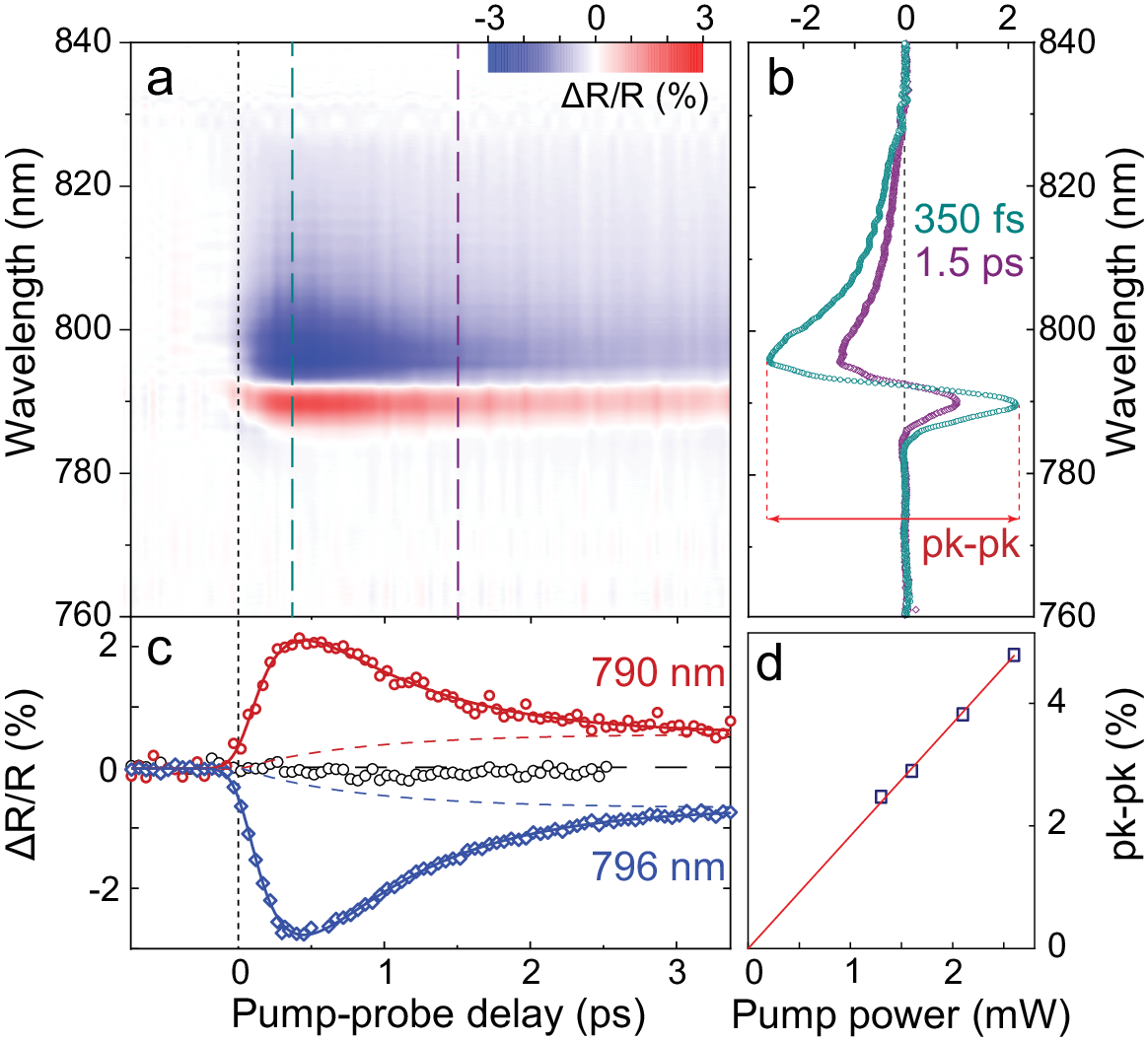}
  \caption{Experimental results. (a) Transient reflectivity variations in the vicinity of the SPP resonance at the Au/air interface ($\lambda\approx793$~nm). Temporal and spectral cross-sections (b,c). The solid lines in (c) are the fitting curves (see text). The black open circles are measured on the flat Au surface (without corrugation). The dashed lines indicate the onset of the thermal effect (as obtained from the fit). (d) Pump power dependence of the peak-to-peak transient reflectivity variations at 350~fs delay (cf. panel b).}
  \label{fig:3}
\end{figure}

The main results of this work demonstrating a transient response of the SPP dispersion to the injection of hot electrons are summarized in Fig.~\ref{fig:3}. Multiple observations can be made from the data presented there. The characteristic operation time well below 1~ps confirms the ultrafast nature of the underlying physical processes. The reflectivity variations of about 2-3\% are registered (Fig.~\ref{fig:3}a), much stronger than those reported previously\cite{AlekhinThesis,TemnovJO2016} where no spectral resolution was introduced. Similar experiments probing the flat, uncorrugated Au surface resulted in $\sim20$ times smaller $\Delta R/R$ variations and no spectral dependence. Additional experiments at reduced laser fluences revealed that observed $\Delta R/R$ remains in the linear regime (Fig.\ref{fig:3}d). The pump-induced dynamics is largely antisymmetric in the spectral domain, indicating the dominant role of a shift of the SPP resonance. This is further corroborated by the cross-sectioned data obtained at $t=350$~fs delay and shown in Fig.~\ref{fig:3}b. 

Further insights can be obtained from the cross-sections in the time domain, as shown in Fig.~\ref{fig:3}c for the two wavelengths exhibiting the strongest reflectivity variations. Owing to the strongly subpicosecond timescales, the contribution from the hot non-thermal electron population needs to be considered, thus ruling out the implementation of the conventional two-temperature models. Attempts to include the hot electron relaxation in a phenomenological (rate equation) way have been introduced \cite{SunPRB94}, and a unified description of the electron dynamics in metals encompassing local and non-local (diffusion) relaxation was recently published \cite{SchiratoNanoph2023}. However, since at spatial scales $\lesssim 100$~nm the hot electron transport is long known to be ballistic (superdiffusive) \cite{BrorsonPRL87,AlekhinPRL17}, the approaches based on the standard equation of diffusion are hardly applicable to our case. Rather, a model based on the six-dimensional Boltzmann equation with realistic electron scattering would be highly desirable. Notably, the time traces in the positive and negative regions of $\Delta R/R$ demonstrate a very high similarity of their dynamical response. A compelling verification consists in a joined fit of these datasets with a logistic-exponential function convoluted with the pump-probe cross-correlation signal $S_{\rm CC}(t)$:

\begin{equation}
    \label{eq:fit}
    \Delta R/R\:(t)=S_{\rm CC}(t)\ast[A(1-e^{-t/\tau_1})e^{-t/\tau_2}+B(1-e^{-t/\tau_2})]
\end{equation}

{\color{red}
Although the function looks identical to that used in Ref.~\cite{delFattiPRB00} with the characteristic times attributed to the electron-electron and electron-phonon relaxation, we emphasize a few important differences. In Ref.~\cite{delFattiPRB00} the laser pulse excited the electronic subsystem in the probed region directly whereas in our case, the perturbation is delivered to the Au surface by way of superdiffusive transport of hot electrons. As such, the rise time of the experimentally registered variations $\Delta R/R$ is governed not by the laser pulse duration, but rather by the emission decay time (estimated to about 250~fs \cite{AlekhinPRL17}) and the electron transport timescale (on the order of $L/v_F \lesssim 100$~fs , where $L = 90$~nm is the Au thickness and $v_F \approx 1.4$~nm/fs is the Fermi velocity \cite{BrorsonPRL87}). The rise time $\tau_1 = 270\pm40$~fs obtained from the fit is in a good agreement with these numbers, corroborating hot electron transport as the main mechanism behind the delivery of the excitation across the Au film. The exact timescale, however, is distorted by the initial distribution of the hot laser-excited electrons in the k-space and the exponentially decaying SPP sensitivity to the depth distribution of the perturbation \cite{RazdolskiPRB19,KuhnePRR2022}. The superdiffusive nature of the non-thermal electrons transport is nonetheless strongly emphasized by the experimentally found delay $t(\Delta R/R|_{\rm max})$ value of $t_{\rm max}\approx350$~fs.

Further, as the response of the local electronic density to the perturbation by hot electrons is operative on intrinsic timescales of the former (on the order of inverse $\omega_p$, i.e. femtoseconds), the decay of the registered signal is intimately related to the dynamics of the non-thermal electronic population. Owing to the ultrafast superdiffusive transport, the hot electrons at delays $>300$~fs are uniformly distributed in the Au film which is why we further consider local relaxation processes. In particular, we largely attribute the longer timescale to the thermalization of the electronic subsystem leading to its subsequent heating. Reducing the frustration of the electron density induced by the non-thermal population, this process in turn gives rise to the slowly evolving thermal contribution (dashed lines in Fig.\ref{fig:3}). The found timescale $\tau_2\approx 840 \pm 55$~fs agrees well with the relaxation times of the non-thermal electron population in Au reported in multiple works \cite{AdPRb92,HohlfeldAPB1997,delFattiPRB00,FannPRB92,SunPRB94,MejardACSPhot2016,KuhnePRR2022}. The picosecond timescale is dominated by the relaxation of the electronic density to the Fermi-Dirac equilibrium albeit at an elevated temperature, giving rise to the slowly evolving thermal contribution (dashed lines in Fig.~\ref{fig:3}c). It is thus can be construed that the first and second terms in Eq.\ref{eq:fit} are dominated by the non-thermal ($A$) and thermal ($B$) electron-induced contributions, respectively. This assignment of the timescales is consistent with our previous work \cite{RazdolskiPRB19}, where two characteristic times of the non-thermal electron relaxation in Au were analyzed. Too short for the hot electron relaxation, the smaller one $\tau_1\approx50$~fs was attributed to the hot electron transport in a 50 nm-thick Au film while the longer time  $\tau_2\approx1$~ps was assigned to the local thermalization of the hot electrons excited by the near-IR laser pulses (photon energy $0.8-1.1$~eV). In our current work, $\tau_1$ is increased by the Fe emitter, while $\tau_2$ is slightly reduced thanks to the higher energy of the laser-excited hot electrons.
}
Moreover, experiments with the varied pump pulse energy clearly revealed a linear dependence of the redshift of the resonance (Fig.~\ref{fig:3}d).

The experimentally observed redshift of $\omega_{\rm SPP}$ indicates ultrafast variations of $\omega_p$ through the transient electron spill-out on the subpicosecond timescale. The response of the native electronic density in Au to the injection of the hot electrons can be primarily described within the time-dependent Feibelman formalism $d_{\perp}(t)$, entailing a transient spatial displacement $\Delta_z n_e(t)$ along the normal to the Au surface. Assuming the Lorentzian shape of the reflectivity dip associated with the SPP excitation (Fig.~\ref{fig:2}b), the transient Feibelman variations of $\omega_p$ can be related to the experimental $\Delta R/R$ data shown in Fig.~\ref{fig:2}. 
{\color{red}
Indeed, for the associated $\omega_p$ variations we can write:
\begin{equation}
    \frac{d\lambda}{d\omega_p}=\frac{\partial \lambda}{\partial k_{\rm SPP}}\cdot\frac{\partial k_{\rm SPP}}{\partial \varepsilon}\cdot\frac{\partial \varepsilon}{\partial \omega_p}
    \label{eq:derivs}
\end{equation}
The terms in the right hand side of this equation can be calculated directly. In particular, because $k_{\rm SPP}=-k_0\sin\theta+2\pi/b$ on our Au grating, where $k_0=2\pi/\lambda$ is the wavenumber and $\theta=45\degree$ is the angle of incidence, we obtain $\partial\lambda/\partial k_{\rm SPP}=b/k_0$. Then, from $k_{\rm SPP}=k_0\sqrt{\varepsilon/(1+\varepsilon)}$ at the air/Au interface we find $\partial k_{\rm SPP}/\partial \varepsilon=0.5\varepsilon^{-1/2}(1+\varepsilon)^{-3/2}$. Recalling that $|\varepsilon|\gg 1$ in this spectral range and neglecting small losses, it simplified to $0.5\varepsilon^{-2}$. Finally, within the Drude model of the dielectric permittivity, we get $\partial\varepsilon/\partial\omega_p=-2\omega_p/\omega^2\approx 2\varepsilon/\omega_p$. Assembling everything together back into Eq.~(\ref{eq:derivs}), we obtain: $\Delta\lambda=b\cdot\delta\omega_p/|\varepsilon|$, where $\delta\omega_p=\Delta\omega_p/\omega_p$ is the relative variation of the plasma frequency.

Further, approximating the SPP resonance with a Lorentzian, we found that experimental reflectivity variations $\Delta R/R|_{\max}-\Delta R/R|_{\min}\approx5\times10^{-2}$ correspond to a resonance wavelength shift $\Delta\lambda\approx0.25$~nm. Combining the last two results together, we find $\delta\omega_p\approx1.2\times10^{-2}$. From here, 
Eq.~(\ref{eq:feibelman}) indicates that, neglecting the small static effects, $\delta\omega_p\approx k_0 d_{\perp}$, and thus an effective value of $d_{\perp}\approx+1.5$~nm at the time delay $t_{\rm max}$ can be retrieved. 
}
For comparison, in the equilibrium, calculations give $d^{(0)}_{\perp}\approx-0.05$~nm at the Au/air interface \cite{RodriguezEcharri21}. Further, the broadening of the SPP resonance is much less pronounced, as it can be inferred from the largely antisymmetric shape in Fig.~\ref{fig:3}b. This indicates the negligibility of the $\mathfrak{Im}\:d_{\perp}$ associated with the surface-induced Landau damping contribution within the Feibelman formalism.

The non-thermal electron injection thus results in huge variations of the spatial distribution of the electronic density in Au on the ultrafast timescale. The $d_{\perp}$ sign change indicates a transition from the spill-in regime characteristic for noble metals\cite{} to the large transient spill-out and associated SPP dispersion shift. Although the effective $d_{\perp}\approx+1.5$~nm largely (more than one order of magnitude) exceeds those computed for the interfaces in the equilibrium\cite{ChristensenPRL17,RodriguezEcharri21}, we emphasize that this is an effective value. As such, within the Feibelman approach, the hot electron-induced correction to the classical, local response approximation by far exceeds that introduced by an interface. Notably, the relative weight of the static, interface contribution can be enhanced in small-scale plasmonic systems. Observing an intriguing interplay of these two mechanisms in essentially quantum plasmonic systems with dimensions below 5-10~nm remains an attractive perspective.

Lastly, we disprove other mechanisms which could be conjectured to contribute to the observed effect. In semiconductor plasmonics, the low density of the free electrons enables its strong modulation by means of injection of hot electrons from the adjacent metal or inducing interband transitions \cite{CaspersOpt2010,GarciaNanoLett2011,HarutyunyanNatNano2015,KriegelJPCL2016,ShcherbakovNatComm2017}. In that method, the transient carrier density ne experiences a sizeable increase, in turn increasing the plasma frequency $\omega_p$ and blueshifting the surface plasmon resonance. We argue that this mechanism can be ruled out in our system for the following reasons. Firstly, the experiments show a consistent {\it redshift} of the SPP resonance which corresponds to the effective reduction of $n_e$. This modification of the electron density upon laser excitation of hot electrons in the Fe emitter is highly unlikely. Secondly, the charge screening in Au will effectively maintain the local charge neutrality. Owing to the large $n_e$ in Au ($6\times10^22$~cm$^{-3}$), this screening is very effective and operative on the timescales given by $\omega_p^{-1}$, i.e., femtoseconds. Conversely, in low-$n_e$ semiconductors, the injected non-equilibrium carriers-induced charge can exhibit long lifetimes, enabling experimental observation of the transient SPP resonance shifts.

Similarly, we emphasize that, owing to the 350~fs timescale, the shift of the SPP dispersion is unrelated to the laser heating of the plasmonic medium \cite{RotenbergOL2008,RotenbergPRB2009,AbbNatComm2014,NovikovNanoLett2020}. Indeed, due to the large Au thickness (much larger than the optical penetration depth, $\sim 20$~nm), there is no direct laser heating at the Au surface. In turn, the phonon heat transport  requires prior electron thermalization and is thus much slower\cite{TemnovNatComm2013,MatternAPL2022}, limited by the speed of sound in Au $v_s\approx 3.2$~nm/ps. Measurements up to 25~ps delays indicated no significant impact of heating at this timescale.

Another recently discussed mechanism \cite{SchiratoNanoph2023} is related to the heating-induced lattice expansion and associated decrease of the electron density. This, in turn, could reduce $\omega_p$ and redshift the SPP resonance. Estimations show that for the laser pulses of about $2~\mu$J energy, in our experimental conditions and assuming a perfect heat contact between Fe and Au, the eventual temperature rise after reaching a thermal equilibrium between the electrons and lattice subsystems can be expected on the order of $100$~K. The volumetric expansion coefficient of about $4\cdot10^{-5}$~K$^{-1}$ in Au (\cite{PamatoJAC2018} and references therein) has been reported, yielding <1\% modulation of the active Au volume, and similarly, the electron density $n_e$. Importantly, this modulation can be achieved at timescales longer than that of the electron-phonon equilibration but shorter than characteristic timescale of the heat transport resulting in the energy flow into the substrate. However, at $350$~fs delays, no electron-phonon equilibration can be expected. Conversely, the lattice remains relatively cold whereas the SPP dispersion has already exhibited a shift (Fig.~\ref{fig:2}), thus ruling out the thermal expansion mechanism.

To summarize, our results bring quantum plasmonic effects onto ultrafast timescales governed by the lifetimes of hot non-thermal electrons. For the first time, we demonstrate the feasibility of controlling plasmon dispersion through transient variations of the spatial electron distribution in commonly available plasmonic metals. Highly energetic non-thermal electrons represent an excellent means for that purpose, featuring fast operation times and large travel distances. The latter allows for the remote laser excitation minimizing the undesirable thermal effects. Strong non-local response of the native electronic density in plasmonic media opens up novel perspectives for engineering transient plasmonic properties of nanostructures with optical stimuli. Introducing ultrashort timescales into quantum plasmonics, our results significantly expand the field of ultrafast nanophotonics.

\begin{acknowledgement}

The authors thank Dr. A.~Melnikov (Martin-Luther-Universität Halle-Wittenberg), Prof. S.~Sanvito (Trinity College Dublin), Dr. P. André D.~Gonçalves (ICFO–Institut  de  Ciencies  Fotoniques) and Prof. Kun Ding (Fudan University) for fruitful discussions.
This work was supported by the National Science Centre Poland (Grant No. DEC-2019/35/B/ST3/00853). V.R. acknowledges the European Regional Development Fund and the State of Brandenburg for the Themis Z TEM (part of the Potsdam Imaging and Spectral Analysis Facility (PISA)).

\end{acknowledgement}

\bibliography{refs}



\end{document}